\documentclass[aps,prl,twocolumn,superscriptaddress,floatfix]{revtex4}

\usepackage{amsmath}
\usepackage{amssymb}
\usepackage[dvips]{graphicx}

\def\non{\nonumber}

\def\mbold#1{\mbox{\boldmath $#1$}}

\def\beq{\begin{equation}}
\def\eeq{\end{equation}}
\def\bea{\begin{eqnarray}}
\def\eea{\end{eqnarray}}
\def\fun#1#2{\lower3.6pt\vbox{\baselineskip0pt\lineskip.9pt
  \ialign{$\mathsurround=0pt#1\hfil##\hfil$\crcr#2\crcr\sim\crcr}}}

\usepackage{ulem}
\def\mbold#1{\mbox{\boldmath $#1$}}

\begin{document}

\title{Borromean Feshbach resonance in $^{11}$Li
   studied via $^{11}$Li($p$,$p'$)}

\author{Takuma Matsumoto}
\email[]{matsumoto@phys.kyushu-u.ac.jp}
\affiliation{Department of Physics, Kyushu University, Fukuoka 819-0395, Japan}

\author{Junki Tanaka}
\affiliation{Research Center for Nuclear Physics, Osaka University,
Ibaraki, Osaka 567-0047, Japan}
\affiliation{Department of Physics, Konan University, Higashinada, Kobe,
Hyogo 658-8501, Japan}
\affiliation{Institut f\"ur Kernphysik, Technische Universit\"at Darmstadt, 64289 Darmstadt, Germany}

\author{Kazuyuki Ogata}
\affiliation{Research Center for Nuclear Physics, Osaka University,
Ibaraki, Osaka 567-0047, Japan}
\affiliation{Department of Physics, Osaka City University, Osaka 558-8585, Japan}

\date{\today}

\begin{abstract}%
A dipole resonance of $^{11}$Li is found by a
$^9$Li + $n$ + $n$ three-body model analysis with the complex-scaling method.
The resonance can be interpreted as a bound state in the $^{10}$Li + $n$
system, that is, a Feshbach resonance in the $^9$Li + $n$ + $n$ system.
As a characteristic feature of the Feshbach resonance of $^{11}$Li,
the $^{10}$Li + $n$ threshold is open above the $^{9}$Li + $n$ + $n$
 one, which reflects a distinctive
property of the Borromean system.
A microscopic four-body reaction calculation for the $^{11}$Li($p$,$p'$)
reaction at 6 MeV/nucleon is performed by taking into account
the resonant and nonresonant continuum states of the three-body
system. The calculation of angular distributions of the elastic and
 inelastic scattering 
 as well as the energy spectrum reproduced a recent experimental
 result. Furthermore, the E1 strength distribution from a Coulomb
 dissociation experiment was also reproduced in this framework. This
 means that the existence of the {\it Borromean Feshbach resonance} is
 consistently answering a longstanding open question of an excited state
 of $^{11}$Li. 
\end{abstract}


\maketitle
\section{Introduction}

Elucidation of resonances, which are omnipresent in different
hierarchies in nature, is one of the most important subjects in physics.
For example, the tetraquark and pentaquark baryons in hadron
physics~\cite{Che16} as well as the so-called Efimov
resonance~\cite{Efi70,Hua14} of ultracold atoms in atomic physics
have attracted the attention of many experimentalists and theorists.
In nuclear physics, various resonances have been discovered and
investigated in detail. Studies of resonances in nuclear physics will
be characterized by diversity.
Nuclei, a self-organized strongly interacting system, show a wide
variety of structures as the atomic number, the mass number,
and the excitation energy change.  From a different
point of view,
we have better knowledge
on the basic interaction that forms many-nucleon systems than in hadron
physics. Various types of resonances, e.g., single-particle resonances,
gas-like $\alpha$ cluster states, and giant resonances have therefore
been investigated on a solid basis.
Nowadays resonant structures for nuclei near and even beyond the
neutron dripline have intensively been proceeded.
Furthermore,  a recent experiment suggested that four neutrons form a
resonance, that is, the so-called tetraneutron~\cite{Kis16}.

Some nuclei near the dripline such
as $^6$He, $^{11}$Li, $^{14}$Be, and $^{22}$C are known as two-neutron
halo nuclei~\cite{Tan85,Suz99,Mor14,Tan10} consisting
of a core nucleus
and two loosely bound neutrons. These nuclei have a Borromean structure,
meaning that there is no bound state for each pair of the three constituents.
Except for $^6$He, experimental information on resonances of such
Borromean nuclei is very scarce. Existence of a resonance of $^{11}$Li, the
firstly discovered Borromean nucleus, is a longstanding open question in
particular~\cite{Iek93,Shi95,Zin97,Nak06,Kor97a,Kor97b,Kan15,Cre02,Gar03,Ers04,Pin12,Hag05,Myo08,Kik13}.
Especially the hadronic scattering and the Coulomb dissociation
experiments provided contradictory results. While the peak structures
were observed with several hadronic scatterings, the low-lying peak of
the recent Coulomb dissociation experiments can be explained by only E1
direct breakup component without resonances and the other apparent
resonant state was not observed\cite{Nak06}. Because of its experimental
difficulties, the hadronic scattering experiments only had less
statistics or poor 
resolutions comparing to Coulomb dissociation experiments.

Very recently,  measurement of the $^{11}$Li($p,p'$) reaction at 6
MeV/nucleon with a high statistic and high resolution has
been performed~\cite{Tan17} to clarify this situation, and a low-lying excited state of $^{11}$Li
has clearly been identified.
In the analysis, the authors adopted a macroscopic model for the
transition of $^{11}$Li combined with the distorted wave Born
approximation (DWBA); a form factor of the isoscalar electric dipole ($E1$)
excitation is assumed.
The macroscopic model, however, does not describe the Borromean
nature of $^{11}$Li and  a microscopic approach to the structure of the
low-lying continuum states of $^{11}$Li is eagerly desired.
On the reaction side, the applicability of DWBA in the energy region of
our interest is quite questionable. In other words, if the reaction
observable {\it suffers} from higher-order processes, it is not trivial at all
to relate the observable and a response of a nucleus to a specific
transition operator. Furthermore, there is no guarantee that a single
operator is responsible for the proton inelastic scattering measured at
backward angles.

The purpose of this paper is to analyze the
$^{11}$Li($p,p'$) cross sections at 6 MeV/nucleon with a sophisticated reaction model,
that is, the microscopic four-body continuum-discretized
coupled-channels method (CDCC)~\cite{Yah12,Mat04,Mat06,Rod08,Mat10}.
A complete set of the $^9$Li + $n$ + $n$ three-body wave
functions in a space relevant to the $^{11}$Li($p,p'$) reaction
is implemented in CDCC and thereby the validity of the continuum structure of $^{11}$Li
is examined.
Classification of the three-body wave
functions with the complex-scaling method (CSM)~\cite{Agu71,Bal71,Aoy06}
suggests a low-lying three-body Feshbach resonance~\cite{Fes58} of
$^{11}$Li, which is the principal finding of the present study.
We discuss also the electric dipole distribution calculated with
the three-body model of $^{11}$Li.

This paper is organized as follows: in Sec. II, we describe the
theoretical framework for the present analysis of the
$^{11}$Li($p,p'$) reaction. In Sec. III, we investigate the resonance of
$^{11}$Li by comparing the theoretical results with the
experimental data. Finally, we give a conclusion in Sec. IV.

\section{Theoretical Framework}
For $^{11}$Li, we  adopt a $^{9}$Li~+~$n$~+~$n$ three-body model,  with
assuming for simplicity that $^9$Li is a spinless and inert particle
that has a na\"{\i}ve shell-model configuration.
This  simplified model has been applied to analyses of some reactions of
$^{11}$Li~\cite{Cre02,Ers04,Pin12}.
Three-body wave functions $\Phi_{I^\pi \nu}$, where $I^\pi$
represents the spin-parity and $\nu$ is the index of eigenenergy,
of $^{11}$Li are obtained by diagonalizing the three-body Hamiltonian:
\begin{eqnarray}
 h&=&K_{r}+K_{y}+V_{nn}+V_{cn}+V_{cn}+V_{cnn}.
\end{eqnarray}
Here, $K_{r}$ and $K_y$ are the kinetic energy operators for the Jacobi
coordinates $\mbold{r}$ and $\mbold{y}$ shown in Fig.~1 in
Ref.~\cite{Mat04}, respectively. $V_{nn}$ ($V_{cn}$) is a two-body
interaction between the two neutrons ($^{9}$Li and a neutron), and
$V_{cnn}$ is a phenomenological three-body force (3BF).
$\Phi_{I^\pi \nu}$ is explicitly antisymmetrized for the exchange between
the two valence neutrons, whereas the exchange between each valence neutron and
a nucleon in $^9$Li is approximately treated by the orthogonality
condition model~\cite{Sai69}.

For understanding properties of the three-body continuum of
$^{11}$Li in more detail, we employ CSM, in which
the radial part of each Jacobi coordinate is transformed as
\begin{eqnarray}
 r\to re^{i\theta_c},\quad
 y\to ye^{i\theta_c}
\end{eqnarray}
with the scaling angle $\theta_c$, and $h$ is rewritten as
$h^{\theta_c}$ accordingly.
As a result of diagonalization of  $h^{\theta_c}$,
eigenstates $\varphi_{I^\pi \nu}^{\theta_c}$ that have complex
eigenenergies $\varepsilon_{I^\pi \nu}^{\theta_c}$ are obtained. A
resonance is identified as
an eigenstate on the complex-energy plane isolated from other nonresonant states;
the real and imaginary parts of the eigenenergy represent the resonant energy
$\varepsilon_{\rm R}$ and a half of the decay width $\Gamma/2$, respectively.
$\varphi_{I^\pi \nu}^{\theta_c}$ are used also in obtaining a
continuous breakup energy spectrum from discrete breakup cross sections
obtained by CDCC, as shown below.

The total wave function $\Psi$ of the $p+{}^{11}$Li reaction system is
obtained by solving the Schr\"{o}dinger equation
\begin{eqnarray}
\left(
K_R +h + \sum_{i \in ^{11}{\rm Li}} v_{0i} +V_{\rm C}
-E\right)\Psi^{(+)}=0,
 \label{Sch}
\end{eqnarray}
where $K_R$ is the kinetic energy operator regarding the coordinate
$\mbold{R}$ between the center-of-mass of $^{11}$Li and $p$. The nuclear
interaction between $p$ and the $i$th nucleon in $^{11}$Li
is denoted by $v_{0i}$. $V_{\rm C}$ is the Coulomb interaction
between $p$ and the center-of-mass of $^{11}$Li; we thus ignore the
Coulomb breakup process.

In CDCC, we assume that the scattering takes place in a model space defined by
\begin{eqnarray}
 {\cal P}&=&\sum_{\gamma}|\Phi_{\gamma}\rangle\langle\Phi_{\gamma}|,
\end{eqnarray}
where $\gamma=(I^\pi,\nu)$. $\Psi^{(+)}$ is then approximated into
\begin{eqnarray}
 \Psi^{(+)} &\approx&
  {\cal P}\Psi^{(+)}=
  \sum_{\gamma}\chi_\gamma^{(+)}(\mbold{R})
  \Phi_{\gamma}\phi_{\rm c}, \label{Eq:Psi}
\end{eqnarray}
where $\chi_\gamma^{(+)}$ is the relative wave
function regarding $\mbold{R}$, and $\phi_c$ is an internal wave
function of $^9$Li.
Inserting Eq.~\eqref{Eq:Psi} into Eq.~\eqref{Sch} leads to a set of
coupled equations for $\chi_\gamma^{(+)}$:
\begin{eqnarray}
&& \left[
  K_R+U_{\gamma\gamma}(\mbold{R})-(E-\varepsilon_\gamma)
 \right]\chi_\gamma^{(+)}(\mbold{R})\non\\
&&\hspace{2cm} =-\sum_{\gamma'\ne\gamma}U_{\gamma\gamma'}(\mbold{R})
 \chi_{\gamma'}^{(+)}(\mbold{R})\label{Eq:cc}
\end{eqnarray}
with $\varepsilon_\gamma=\langle\Phi_\gamma|h|\Phi_\gamma\rangle$.
This is called CDCC equations, which is solved under the standard
boundary condition~\cite{Yah12}.
In the microscopic four-body CDCC, coupling potentials between
$\Phi_\gamma$ and $\Phi_{\gamma'}$, $U_{\gamma\gamma'}$, are obtained by
\begin{eqnarray}
 U_{\gamma\gamma'}(\mbold{R})=
  \int d\mbold{s} \rho_{\gamma\gamma'}(\mbold{s})
  {v}_{0i}(\rho,E,|\mbold{R}-\mbold{s}|)
  +\frac{3e^2}{R}\delta_{\gamma\gamma'},
  \label{rhogg}
\end{eqnarray}
where the transition densities are defined by
\begin{eqnarray}
 \rho_{\gamma\gamma'}(\mbold{s})&=&
  \langle\Phi_\gamma\phi_{\rm c}|\sum_{j=1}^{11}\delta(\mbold{s}-\mbold{s}_j)
  |\Phi_\gamma'\phi_{\rm c}\rangle.
\end{eqnarray}
Here $\mbold{s}_j$ is the coordinate of the $j$th nucleon in $^{11}$Li
relative to the center-of-mass of $^{11}$Li.

By solving Eq.~(\ref{Eq:cc}), one obtains a transition matrix element
$T_\gamma$ from which a cross section to the ground state or
a discretized-continuum state of $^{11}$Li can be evaluated.
To obtain a continuous breakup energy spectrum, we employ the
smoothing method based on CSM proposed in Ref.~\cite{Mat10}.
Consequently, the double differential breakup cross section
with respect to the energy $\varepsilon$ of the $^{9}$Li~+~$n$~+~$n$ system
measured from the three-body threshold and the solid angle $\Omega$
of the center-of-mass of the three particles
is obtained by
\begin{eqnarray}
\frac{d^2 \sigma}{d\varepsilon d\Omega}
=
\frac{1}{\pi}
{\rm Im}
\sum_{\gamma'}
\frac{T_{\gamma'}^{\theta_c} \tilde{T}_{\gamma'}^{\theta_c} }
{\varepsilon-\varepsilon_{\gamma'}^{\theta_c}},
\label{ddbux}
\end{eqnarray}
where
\begin{eqnarray}
\tilde{T}_{\gamma'}^{\theta_c}=
\sum_\gamma \left\langle \tilde{\varphi}_{\gamma'}^{\theta_c}
\Big|
C(\theta_c)
\Big|
\Phi_{\gamma}
\right\rangle
T_\gamma,
\label{csm-T1}
\end{eqnarray}
\begin{eqnarray}
T_{\gamma'}^{\theta_c}=
\sum_\gamma
T_\gamma^*
\left\langle
\Phi_{\gamma}
\Big|
C^{-1}(\theta_c)
\Big|
\varphi_{\gamma'}^{\theta_c}
\right\rangle
\label{csm-T2}
\end{eqnarray}
with $C(\theta_c)$ and $C^{-1}(\theta_c)$ being the complex-scaling
operator and its inverse, respectively.
As shown in Eq.~(\ref{ddbux}), the breakup energy spectrum is given by an
incoherent sum of the contributions from the eigenstates of
$h^{\theta_c}$. This property is crucial to clarify the role of
a resonance in describing breakup observables.

\section{Results and Discussion}

\subsection{Numerical inputs}

We take the Minnesota force~\cite{Tho77}
for $V_{nn}$ and the interaction used in Ref.~\cite{Esb92} is adopted
as $V_{cn}$. The $V_{cn}$ generates a resonance of $^{10}$Li in the $0p_{1/2}$
state with the resonant energy (decay width) of $0.46$~MeV ($0.36$~MeV).
This resonance is denoted by $^{10}$Li below for simplicity.
This value of the resonant energy is in good agreement with
the latest experimental data~\cite{Cav17}. For $V_{cnn}$, we adopt the
volume-type 3BF~\cite{Mat14} given by a product of Gaussian functions for
the two Jacobi coordinates; the range parameter for each coordinate is
set to 2.64 fm and the strength $V_3$ is
determined to optimize the ground state energy
$\varepsilon_0=-0.369$~MeV~\cite{Smi08} of $^{11}$Li.
We employ the Jeukenne-Lejeune-Mahaux (JLM) effective nucleon-nucleon
interaction~\cite{Jeu77} as $v_{0i}$.
As in the preceding works~\cite{Mat11,Ich12,Guo13},
a normalization factor $N_I$ for the
imaginary part of the JLM interaction is introduced; $N_I$ is determined to
be 0.55 so as to reproduce both the elastic and breakup cross section
data around 100$^\circ$, where the breakup cross section data exist.
Note that we do not include any other adjustable parameters.
Eigenstates of $h$ and $h^{\theta_c}$ are obtained by the Gaussian
expansion method~\cite{Hiy03}, where we adopt the parameter
set II for $h$  and set III for $h^\theta_c$ shown in Table~I in Ref.~\cite{Mat10}.
In CSM, the scaling angle $\theta_c$ is set to $20^\circ$.
In CDCC calculation, we select the $\Phi_{I^\pi\nu}$ with
$\varepsilon<5$ MeV and the resulting number of states is 93, 111, and
131 for $I^\pi=0^+$, $1^-$, and $2^+$, respectively. The model space
gives good convergence of the elastic and breakup cross sections.

\subsection{Structure of the $1^-$ continuum}

\begin{figure}[htbp]
\begin{center}
\includegraphics[width=0.45\textwidth,clip]{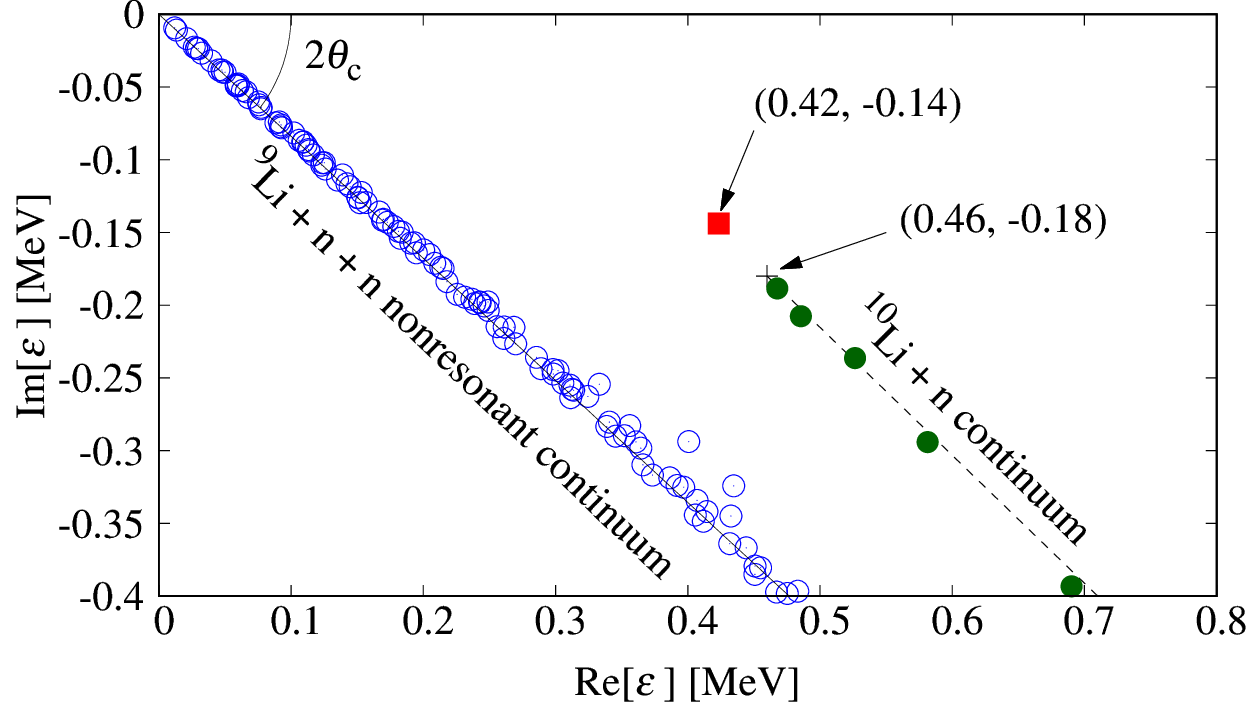}
\caption{Eigenenergies for $1^-$ states calculated with CSM on the
 complex-energy plane measured from the $^9$Li~+~$n$~+~$n$
 threshold. The scaling angle $\theta_{c}$ is taken to be
20$^\circ$, and the
 cross mark shows the $^{10}$Li-$n$ threshold on the complex plane.}
\label{fig:csm}
\end{center}
\end{figure}

In Fig.~\ref{fig:csm} we plot the eigenenergies of $h^{\theta_c}$
with $I^\pi=1^{-}$ on the complex-energy plane.
The solid square shows the three-body
resonance of $^{11}$Li having $\varepsilon_{\rm R}=0.42$ MeV and
$\Gamma/2=0.14$ MeV.
The open circles represent three-body nonresonant continuum states
of the $^9$Li + $n$ + $n$  system,
whereas the closed circles indicate two-body continuum
states between the valence neutron and $^{10}$Li.
One finds that the three-body resonance is located near the
$^{10}$Li-$n$ threshold and the energy of the valence neutron is negative
with respect to $^{10}$Li. This indicates that the dipole resonance of
$^{11}$Li is a Feshbach resonance~\cite{Fes58} in a three-body system
as discussed below.

To clarify the property of the $^{11}$Li continuum states in more detail,
we evaluate an overlap defined by
\begin{eqnarray}
 \alpha_\nu^{\theta_c}&=&2\langle
  \tilde{\varphi}_{1^- \nu}^{\theta_c}
  |{\phi}_{\frac{1}{2}^-}^{\theta_c}\rangle\langle
  \tilde{\phi}_{\frac{1}{2}^-}^{\theta_c}|
  \varphi_{1^- \nu}^{\theta_c}\rangle,
\end{eqnarray}
where $\phi_{\frac{1}{2}^-}^{\theta_c}$ is the complex-scaled wave
function of $^{10}$Li. The factor of 2 means the existence of two
pairs of the $^9$L+$n$ system in $^{11}$Li.
In general, $\alpha_{\nu}^{\theta_c}$ becomes complex, and its real part
can be interpreted as a probability~\cite{Kur07}, in this case,
the probability that the $\nu$th $1^-$ state of $^{11}$Li contains
$^{10}$Li.

For the $^9$Li~+~$n$~+~$n$ nonresonant continuum states
(the open circles in Fig.~\ref{fig:csm}),
which are expected not to contain $^{10}$Li, the real part of
$\alpha_{\nu}^{\theta_c}$ is found to be almost 0.
On the other hand, the real part of $\alpha_{\nu}^{\theta_c}$
for the $^{10}$Li~+~$n$ continuum states (the closed circles)
is larger than 0.9. These results suggest that the real part of
$\alpha_{\nu}^{\theta_c}$ is a good measure for the existing probability
of $^{10}$Li in the $1^-$ continuum states of $^{11}$Li.
It is found that the real part of $\alpha_{\nu}^{\theta_c}$
for the $^{11}$Li resonance (the solid square) is 0.92, meaning
that this state has a similar structure to that of the
$^{10}$Li~+~$n$ continuum states. Because
$\varepsilon_{\rm R}$ of $^{10}$Li is higher than
$\varepsilon_{\rm R}$ of the $^{11}$Li resonance, one can interpret
the $^{11}$Li resonance as a {\lq\lq}bound state'' of the $^{10}$Li~+~$n$
system, that is, a Feshbach resonance~\cite{Fes58}.

\begin{figure}[htbp]
\begin{center}
\includegraphics[width=0.45\textwidth,clip]{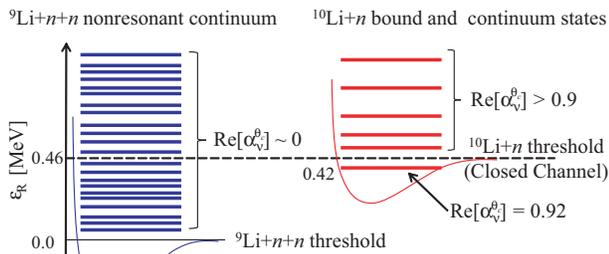}
\caption{A schematic representation of complex-scaled states of
 $^{11}$Li for $I^\pi=1^-$ is shown.}
\label{fig:level}
\end{center}
\end{figure}

In Fig.~\ref{fig:level}, we summarize properties of
the complex-scaled states shown in Fig.~\ref{fig:csm}. In the three-body
Feshbach resonance, the
$^9$Li + $n$ + $n$ threshold energy is lower than the
$^{10}$Li-$n$ threshold, which is a distinctive
character of the Borromean system. We thus refer to this resonance as
{\it a Borromean Feshbach resonance}.
It should be noted that some indications of a dipole resonance
in $^{11}$Li have been discussed in preceding studies~\cite{Pin12,Gar03}.
It will be interesting to see the correspondence between these findings
and the result in the current study.
In the following subsections, we discuss how the $1^-$ resonance
appears in reaction observables.

\subsection{$^{11}$Li resonance in proton inelastic scattering}

\begin{figure}[htbp]
\begin{center}
\includegraphics[width=0.45\textwidth,clip]{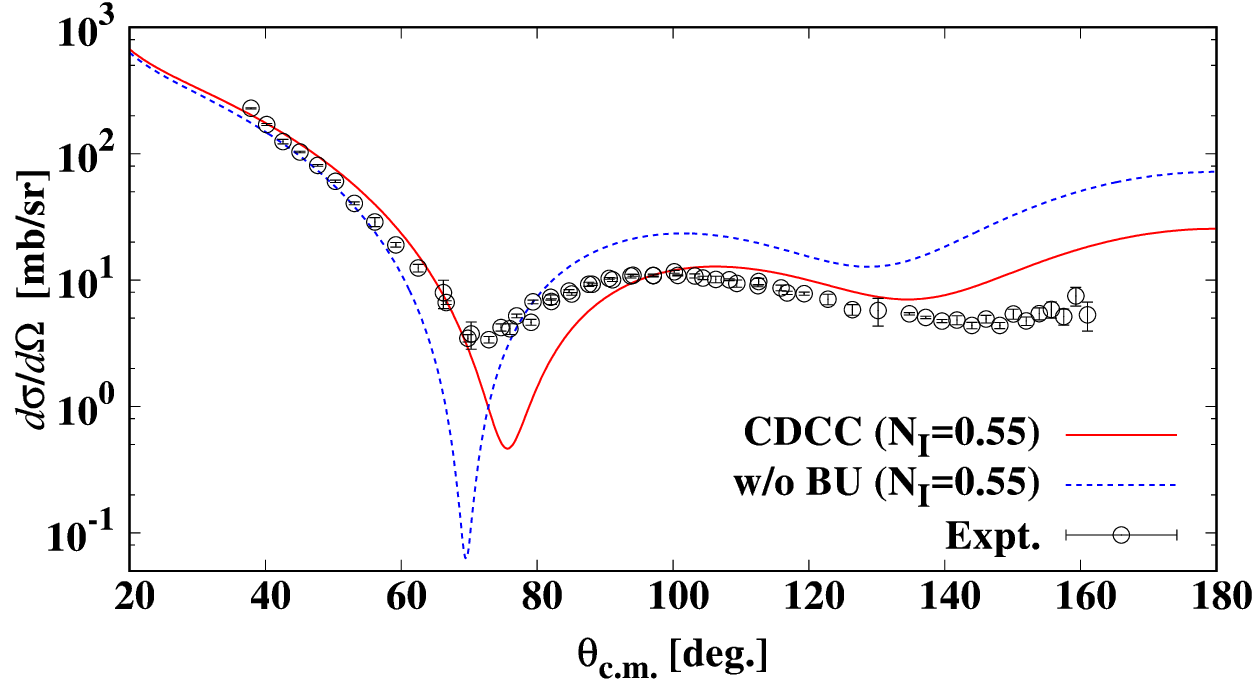}
\caption{Angular distribution of the differential elastic cross section
 for the $^{11}$Li~+~$p$
 scattering at 6 MeV/nucleon~\cite{Tan17}. The solid and
 dotted lines represent results of the microscopic
 four-body CDCC and without breakup channels, respectively.}
\label{fig:elastic}
\end{center}
\end{figure}

First, we discuss the proton elastic scattering on $^{11}$Li that is
used as a primary constraint for the adjustable parameter $N_I$
contained in the present reaction model.
Figure~\ref{fig:elastic} shows the
angular distribution of the elastic cross section at 6~MeV/nucleon.
The solid line is the result of the microscopic four-body CDCC
calculation; $N_I=0.55$ is chosen so as to reproduce the data
around 100$^\circ$, in which the proton inelastic scattering data are
measured.
We remark here that
because the $N_I$ dependence of the elastic cross section is not
very strong, for the fine tuning of $N_I$, we have used also the data for
the inelastic scattering shown below. 
The solid line agrees well with the data also at forward angles.
One may notice a deviation of the result from the data around the
dip of the cross section. It is known that in this region
a spin-orbit part of the distorting potential, which is disregarded in the
present study, plays an important role.
It should be noted also that, to be strict,
the JLM is applicable to nucleon scattering above 10 MeV~\cite{Jeu77}.
Considering these things, we conclude that the agreement
between the solid line and the experimental data is satisfactory.
The dotted line is the result with neglecting the breakup channels
of $^{11}$Li. One sees that the breakup effect represented by
the difference between the dotted and solid lines is significant
for the elastic scattering.

\begin{figure}[htbp]
\begin{center}
\includegraphics[width=0.45\textwidth,clip]{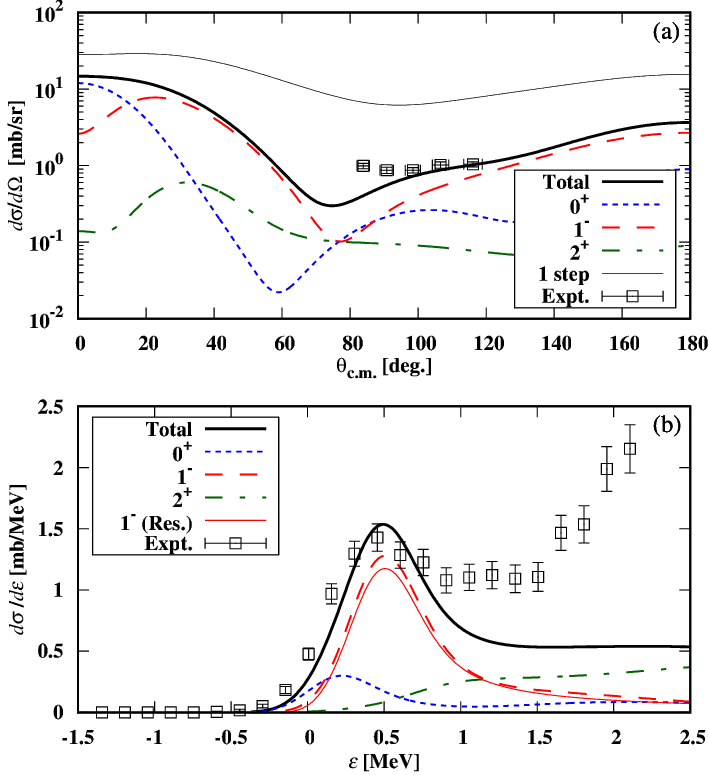}
\caption{(a) Angular distribution of the differential breakup cross
 section and (b) the breakup cross section as a function of the
 three-body breakup energy $\varepsilon$ of $^{11}$Li in the
 $^{11}$Li($p,p'$) reaction~\cite{Tan17}.
The cross section in (a) is obtained by integrating
 $d^2\sigma/(d\varepsilon d\Omega)$ over $\epsilon$ from 0 MeV to 1.13
 MeV, and that in (b) over $\theta_{\rm c.m.}$ from 115$^\circ$ to
 124$^{\circ}$.
 The dotted,  dashed, and dot-dashed lines represent calculated cross
 sections to the $0^+$,  $1^-$, and $2^+$ breakup states, respectively,
 and the sum of them is shown by the thick solid line. The thin solid
 line shows the total cross section calculated with a one-step
 approximation in panel (a) and the contribution of the three-body
 resonance of $^{11}$Li in panel (b).}
\label{fig:brxsec}
\end{center}
\end{figure}

In Fig.~\ref{fig:brxsec}(a) we
show the angular distribution of the breakup cross section;
$d^2\sigma/(d\varepsilon d\Omega)$
is integrated over $\varepsilon$ from 0~MeV to 1.13~MeV
so as to cover well the peak structure of the cross section in Fig.~\ref{fig:brxsec}(b).
The thick solid line represents the result
of the microscopic four-body CDCC; it reproduces the
experimental data around 100$^\circ$, as $N_I$ is chosen so.
The slight deviation of the solid line from the data around
80$^\circ$ will come from the same reason as for the elastic
cross section.
The dotted, dashed and dot-dashed lines
represent the breakup cross sections to the $0^+$, $1^-$ and $2^+$ states,
respectively. One sees that the breakup cross section to the $1^-$
state is dominant
but the $0^+$ and $2^+$ components are
not negligible in the region where the experimental
data exist.
In other words, a model that assumes a pure dipole transition of $^{11}$Li
will not explain the measured cross sections unless an unrealistic
structural model of $^{11}$Li is adopted. Furthermore, since the
transition potential adopted in the
present calculation cannot be written as a simple functional
form, to use a single transition operator can not be justified.
Our final remark on Fig.~\ref{fig:brxsec}(a) is the importance of the
coupled-channel effects.
The thin solid line shows the result of a one-step calculation that
severely overestimates the  thick
solid line by about one-order at middle angles.
We therefore conclude that DWBA is not applicable to the $^{11}$Li($p,p'$) at
6 MeV/nucleon.

In Fig.~\ref{fig:brxsec}(b), we show the breakup cross section with respect
to the three-body energy $\varepsilon$ after breakup, which is obtained
by integrating $d^2 \sigma/(d\varepsilon d\Omega)$ over $\theta_{\rm c.m.}$ from
115$^\circ$ to 124$^\circ$.
Here, we have taken into account the energy resolution of the
experimental data. The total breakup cross section represented by the
thick solid line reproduces the experimental data up to
$\varepsilon\sim1.0$ MeV including a low-lying peak. One sees that the
contribution from the dipole resonance of $^{11}$Li shown by the thin
solid line dominates the low-lying peak. 
It should be noted that the
peak position (energy) of the experimental data as well as the thin
solid line is somewhat higher than the resonant energy,
$\varepsilon_{\rm R}=0.42$~MeV, which is mainly due to the energy
resolution.   
Furthermore, the calculated resonant width $\Gamma=0.28$ MeV
is narrow compared with the evaluation $\Gamma=1.15\pm0.06$ MeV in
Ref.~\cite{Tan17}. This is because of the contributions
from the $0^+$ and $2^+$ nonresonant components.
It can be concluded, therefore, that the nonresonant
components should be properly evaluated and subtracted from the measured
spectrum to extract reliable information on the width of the resonance.
The calculated cross section undershoots the data for
$\varepsilon\gtrsim 1.0$ MeV, which will be due to some other degrees of
freedom that are not taken into account in the present calculation, for
example, a transition to higher spin states and a core excitation in
$^9$Li.
Thus, we have shown through CDCC calculation that the three-body
structure of $^{11}$Li both in the bound and continuum states including
the dipole resonance of $^{11}$Li shown in Fig.~\ref{fig:csm} is
consistent with the measured cross sections.

\begin{figure}[htbp]
\begin{center}
\includegraphics[width=0.45\textwidth,clip]{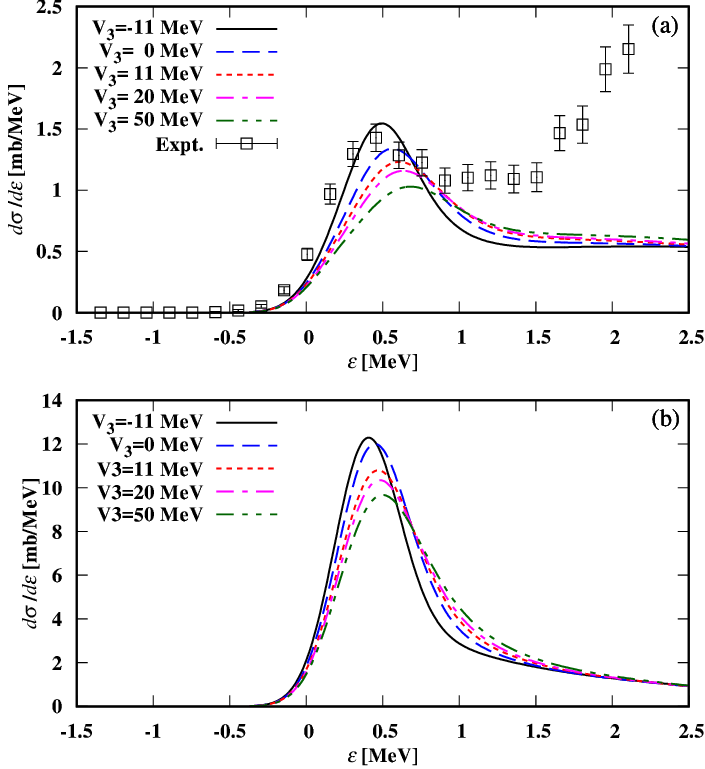}
\caption{The same as Fig.~\ref{fig:brxsec}(b) but for the dependence of 
the strength $V_3$ of the 3BF for the $1^-$ states. Panels (a) and (b)
correspond to the result with CDCC and the one-step calculation, respectively.
The solid, dashed, dotted, dash-dotted, dash-dot-dotted lines show the
 results with $V_3=-11$, 0, 11, 20, and 50~MeV, respectively. }
\label{fig:brxv3}
\end{center}
\end{figure}

Next, we investigate the behavior of the breakup energy
spectrum of $^{11}$Li with varying the strength $V_3$ of the 3BF for
the $1^-$ states. For the $0^+$ and $2^+$ states, $V_3$ is kept being
$-11$~MeV that is determined to reproduce the ground state energy of
$^{11}$Li.
In Ref.~\cite{Gar03}, a certain correspondence between the $^{11}$Li
resonance and the $^{10}$Li resonance has been investigated by changing
the $^9$Li-$n$ interaction. 
We do not change the $^{9}$Li-$n$ interaction, however, in this work to
keep the physical resonant energy of $^{10}$Li and the binding energy of 
$^{11}$Li. 
Figure~\ref{fig:brxv3}(a) shows the breakup energy spectrum of $^{11}$Li.
The solid line is the same as in Fig.~\ref{fig:brxsec}(b). The dashed,
dotted, dash-dotted, dash-dot-dotted lines show the results with
$V_3=0$, 11, 20, and 50~MeV, respectively. It is found that
for $V_3 \ge 11$~MeV, the dipole resonance disappears and its
contribution to the cross section is dispersed into those from
other $^{10}$Li-$n$ nonresonant states. Consequently, even when the
dipole resonance does not exist, the breakup cross section has a peak
at a certain $\varepsilon$ depending on $V_3$. However, the consistency with
the measured cross section is obtained only when $V_3=-11$~MeV, that is,
with the dipole resonance shown in Fig.~\ref{fig:csm}.

It will be interesting to do this analysis with the one-step calculation.
As shown in Fig.~\ref{fig:brxv3}(b), the dependence of the cross section on
$V_3$ is somewhat weakened. One may conclude, therefore, that the
higher-order coupling emphasizes the $V_3$ dependence. This can be an
advantage for the low-energy proton inelastic scattering to clarify the
existence of the dipole resonance in $^{11}$Li.

\subsection{Electric dipole transition}

\begin{figure}[htbp]
\begin{center}
\includegraphics[width=0.45\textwidth,clip]{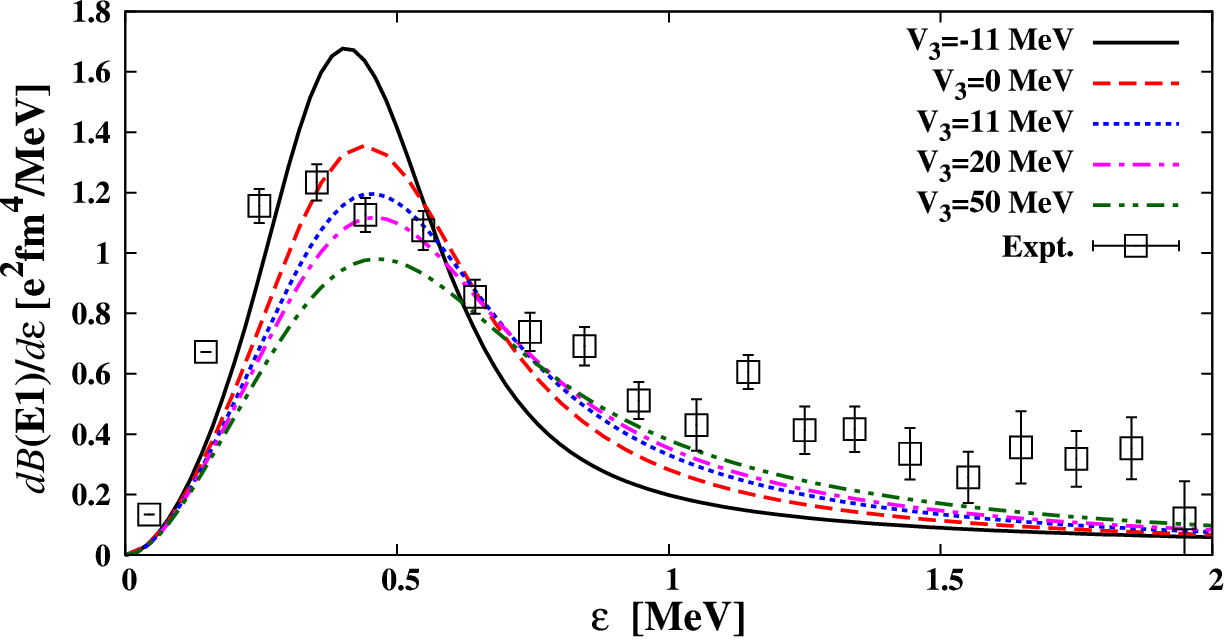}
\caption{$E1$ strength distribution as a function of the
 internal energy of $^{11}$Li. 
 The calculated results are taken into account the experimental energy 
 resolution, and the correspondence between each
 line and $V_3$ is the same as in Fig.~\ref{fig:brxv3}.
 The experimental data are taken from  Ref.~\cite{Nak06}.}
 \label{fig:BE1}
\end{center}
\end{figure}

The dipole resonance has been discussed through the observation of
the electric dipole distribution, $dB({\rm E1})/d\varepsilon$.
In Fig.~\ref{fig:BE1}, we show $dB({\rm E1})/d\varepsilon$
calculated with the present three-body model of $^{11}$Li and
the experimental data~\cite{Nak06}.
The correspondence between each
line and $V_3$ is the same as in Fig.~\ref{fig:brxv3}.
For the theoretical calculation, the continuous energy spectrum of the
$E1$ strength can be obtained by using the same procedure described in
Eqs.~\eqref{ddbux}-\eqref{csm-T2}, in which $T_\gamma$ is replaced by the
$E1$ transition matrix element.
We have taken into account the energy resolution of the
experimental data.

The peak of the solid and dashed lines are at 0.42~MeV and 0.44~MeV,
respectively, reflecting the eigenenergy of the dipole resonance
corresponding to $V_3$. Once the dipole resonance disappears, the
peak position has no dependence on $V_3$ and is fixed at 0.46~MeV,
the threshold energy of the $^{10}$Li-$n$ channel. Because
the resonant energy is very close to the $^{10}$Li-$n$ threshold,
it will be difficult to draw a conclusion from Fig.~\ref{fig:BE1}
on the existence of the dipole resonance.
Another remark is that the experimental data for
$dB({\rm E1})/d\varepsilon$ are extracted from
a breakup cross section of $^{11}$Li by a $^{208}$Pb target at
70~MeV/nucleon. Even though the events corresponding to forward
scattering are selected, the nuclear breakup components and
higher-order effect still may play a role. Furthermore, at
larger $\varepsilon$, contributions from target excitation can
be expected, as indicated by a recent analysis of $^6$He breakup
by $^{208}$Pb at 70~MeV/nucleon~\cite{Sun19}. Taking these
into account, the calculated results describe the experimental
data semi-quantitatively. Although the behavior of the data at small
$\varepsilon$ seems to be reproduced best by the solid line,
it will be not so conclusive in the current situation.

\section{Conclusion}
In conclusion, we have found a dipole resonance in $^{11}$Li at
$\varepsilon_{\rm R}=0.42$~MeV with the width $\Gamma$ of 0.28~MeV
in a $^9$Li + $n$ + $n$ three-body
model calculation with CSM. The continuum structure of the three-body
system including the resonance has been validated by the good
agreement between the results of the microscopic four-body CDCC
calculation and the recently measured $^{11}$Li($p,p'$) data at
6~MeV/nucleon for both the angular distribution and the breakup energy
spectrum. Important remarks on the comparison with the experimental data
are i) contribution of not only the resonance but also the nonresonant
continuum states are important, ii) a one-step calculation (DWBA) does
not work at all, and iii) the transition operator cannot be written in a
simple form as assumed in preceding studies.
It is also found that the present three-body model of
$^{11}$Li can reproduce qualitatively the $E1$ strength distribution.
The peak of the breakup energy distribution of the $(p,p')$ process
turns out to reflect the behavior of the dipole resonance, and
the data are explained well when a resonance with
$\varepsilon_{\rm R}=0.42$ and $\Gamma=0.28$~MeV exists.
On the other hand, the correspondence between the $E1$ strength
distribution and the dipole resonance is found to be less clear.

The structure-reaction combined analysis carried out in this work
will bring non-contradictory understandings of the hadronic scattering
and the Coulomb dissociation experiments.
The $^{11}$Li resonance is interpreted as a bound state of
the valence neutron with respect to $^{10}$Li, that is, a  Borromean Feshbach
resonance. It should be noted that the $^{10}$Li-$n$ threshold is
 above the $^{9}$Li + $n$ + $n$ three-body threshold,
which is a distinctive
character of a Borromean system.
The ordinary Feshbach resonance has intensively been discussed mainly in atomic
physics. The finding of the Borromean Feshbach resonance in the present
study will be characterized by its appearance in a Borromean system that
is unique in the nucleonic system.
Another important feature is that we have some pieces of information on
the interactions between the constituents of $^{11}$Li. This allows one to
carry out realistic studies on the $^{11}$Li resonance. Nevertheless,
more information on the $n$-$^9$Li interaction will be desired to make
our understanding of the continuum structure of $^{11}$Li more profound
and complete. Inclusion of the intrinsic spin of $^{9}$Li as well as the
excitation of the $^{9}$Li core will also be very important.

\section*{Acknowledgments}

The authors are grateful to Prof. I.~Tanihata and Prof. N.~Aoi for
fruitful discussions.
J.T. gratefully acknowledges the support by the Hirao Taro Foundation
of the Konan University Association for Academic Research.
This work has been supported in part by Grants-in-Aid
of the Japan Society for the Promotion of Science (Grants
Nos. JP16K05352 and JP18K03650).


\begin{thebibliography}{9}

\bibitem{Che16}
 H. X. Chen {\it et al.}, Phys. Rep. {\bf 639}, 1 (2016).

\bibitem{Efi70}
 V. Efimov, Phys. Lett. B {\bf 33}, 563 (1970).

\bibitem{Hua14}
B. Huang {\it et al.}, Phys. Rev. Lett. {\bf 112}, 190401 (2014).

\bibitem{Kis16}
 K. Kisamori {\it et al.}, Phys. Rev. Lett. {\bf 116}, 052501 (2016).

\bibitem{Tan85}
I. Tanihata {\it et al.}, Phys. Rev. Lett. {\bf 55}, 2676 (1985);
Phys. Lett. B {\bf 206}, 592 (1988).

\bibitem{Suz99}
T. Suzuki {\it et al}.,
Nucl. Phys. A {\bf 658}, 313 (1999).

\bibitem{Mor14}
T. Moriguchi {\it et al}.,
Nucl. Phys. A {\bf 929}, 83 (2014).

\bibitem{Tan10}
K.~Tanaka \textit{et~al.},
\newblock Phys. Rev. Lett. {\bf 104}, 062701 (2010).

 \bibitem{Iek93}
    K. Ieki {\it et al.}, Phys. Rev. Lett. {\bf 70}, 730 (1993);
    D. Sackett {\it et al.}, Phys. Rev. C {\bf 48}, 118 (1993).

 \bibitem{Shi95}
    S. Shimoura {\it et al.}, Phys. Lett. B {\bf 348}, 29 (1995).

 \bibitem{Zin97}
   M. Zinser {\it et al}., Nucl. Phys. A {\bf 619}, 151 (1997).

 \bibitem{Nak06}
   T. Nakamura {\it et al.},
   Phys. Rev. Lett. {\bf 96}, 252502 (2006).

 \bibitem{Kor97a}
   A. A. Korsheninnikov {\it et al.},
   Phys. Rev. C {\bf 53}, R537 (1997).

 \bibitem{Kor97b}
   A. A. Korsheninnikov {\it et al.},
   Phys. Rev. Lett. {\bf 78}, 2317 (1997).

 \bibitem{Cre02}
   R. Crespo {\it et al.},
   Phys. Rev. C {\bf 66}, 021002(R) (2002).

 \bibitem{Ers04}
   S. N. Ershov {\it et al.},
   Phys. Rev. C {\bf 70}, 054608 (2004).

 \bibitem{Pin12}
   E. C. Pinilla {\it et al.},
   Phys. Rev. C {\bf 85}, 054610 (2012).

 \bibitem{Gar03}
   E. Garrido {\it et al.},
   Nucl. Phys. A {\bf 722}, 221c (2003).

 \bibitem{Hag05}
   K. Hagino and H. Sagawa, Phys. Rev. C {\bf 72}, 044321 (2005).

 \bibitem{Myo08}
   T. Myo {\it et al.},
   Prog. Theor. Phys. {\bf 119}, 561 (2008).

 \bibitem{Kik13}
   Y. Kikuchi {\it et al.},
   Phys. Rev. C {\bf 87}, 034606 (2013).

 \bibitem{Kan15}
   R. Kanungo {\it et al.},
   Phys. Rev. Lett. {\bf 114}, 192502 (2015).

 \bibitem{Tan17}
   J. Tanaka {\it et al.}, 
   Phys. Lett. B {\bf 774}, 268 (2017).

\bibitem{Yah12}
  M.~Yahiro {\it et al.},
  Prog. Theor. Exp. Phys. {\bf 2012}, 01A206 (2012).

\bibitem{Mat04}
  T.~Matsumoto, \textit{et al}.,
  Phys.\ Rev.\ C {\bf 70}, 061601(R) (2004).

\bibitem{Mat06}
  T.~Matsumoto, \textit{et al}.,
  Phys.\ Rev.\ C {\bf 73}, 051602(R) (2006).

\bibitem{Rod08}
  M.~Rodr\'{i}guez-Gallardo, \textit{et al}.,
  \newblock
  Phys.\ Rev.\ C {\bf 77}, 064609 (2008).

\bibitem{Mat10}
  T.~Matsumoto {\it et al.},
  Phys.\ Rev.\  C {\bf 82}, 051602 (2010).

 \bibitem{Agu71} J. Aguilar and J.M. Combes,
   Commun.~Math.~Phys. {\bf 22}, 269 (1971).

 \bibitem{Bal71} E. Balslev and J.M. Combes,
   Commun.~Math.~Phys. {\bf 22}, 280 (1971).

 \bibitem{Aoy06}
  S.~Aoyama {\it et al.},
  Prog.\ Theor.\ Phys.\ {\bf 116}, 1 (2006).

\bibitem{Fes58}
  H. Feshbach, Ann. Phys. (N.Y.) {\bf 5}, 357 (1958); {\bf 19}, 287 (1962).

\bibitem{Sai69}
  S.~Saito,
  \newblock Prog. Theor. Phys. {\bf 41}, 705, (1969).

\bibitem{Tho77}
  D.~R.~Thompson {\it et al.},
  Nucl. Phys. A {\bf 286}, 53 (1977).

 \bibitem{Esb92}
    H. Esbensen and G. F. Bertsch,
   Nucl. Phys. A {\bf 542}, 310 (1992).

 \bibitem{Cav17}
   M. Cavallaro {\it et al.},
   Phys. Rev. Lett. {\bf 118}, 012701 (2017).

 \bibitem{Mat14}
   T. Matsumoto and M. Yahiro,
   Phys. Rev. C {\bf 90}, 041602(R) (2014).

 \bibitem{Smi08}
   M. Smith {\it et al.},
   Phys. Rev. Lett. {\bf 101}, 202501 (2008).


\bibitem{Jeu77}
  J.-P. Jeukenne {\it et al.},
  Phys. Rev. C \textbf{16}, 80 (1977).

\bibitem{Mat11}
  T. Matsumoto {\it et al.},
  Phys. Rev. C {\bf 83}, 064611 (2011).

\bibitem{Ich12}
  D. Ichinkhorloo
  Phys. Rev. C {\bf 86}, 064604 (2012).

\bibitem{Guo13}
  H. Guo {\it et al.},
  Phys. Rev. C {\bf 87}, 024610 (2013).

\bibitem{Hiy03}
  E.~Hiyama {\it et al.},
  Prog.\ Part.\ Nucl.\ Phys.\ {\bf 51}, 223 (2003).

\bibitem{Kur07}
  C. Kurokawa and K. Kat\=o, Nucl. Phys. A {\bf 792}, 87 (2007).

\bibitem{Sun19}
Y. Sun {\it et al.}, in preparation.
\end{thebibliography}
\end{document}